\documentclass[aps,pre,twocolumn,showpacs,groupeaddress]{revtex4-1}
\usepackage{graphicx}
\usepackage{dcolumn}
\usepackage{bm}

\begin{document}
\title{Critical behavior of a quantum chain with four-spin interactions in the presence 
of longitudinal and transverse magnetic fields}

\author{B. \surname{Boechat}}
\email{bmbp@if.uff.br}
\affiliation{Departamento de F\'{\i}sica, Universidade Federal Fluminense\\
Av. Litor\^anea s/n,  Niter\'oi, 24210-340, RJ, Brazil}
\author{O. F. \surname{de Alcantara Bonfim}}
\email{bonfim@up.edu}
\affiliation{Department of Physics, University of Portland, Portland, Oregon 97203, USA}
\author{J. \surname{Florencio}}
\email{jfj@if.uff.br}
\affiliation{Departamento de F\'{\i}sica, Universidade Federal Fluminense\\
Av. Litor\^anea s/n,  Niter\'oi, 24210-340, RJ, Brazil}
\author{A. \surname{Saguia}}
\email{amen@if.uff.br}
\affiliation{Departamento de F\'{\i}sica, Universidade Federal Fluminense\\
Av. Litor\^anea s/n,  Niter\'oi, 24210-340, RJ, Brazil}

\date{\today}

\begin{abstract}

We study the ground-state properties of a spin-1/2 model on a chain containing four-spin 
Ising-like interactions in the presence of both transverse and longitudinal magnetic fields. 
We use entanglement entropy and finite-size scaling methods to obtain 
the phase diagrams of the model.  Our numerical calculations reveal a rich variety 
of phases and the existence of multi-critical points in the system. We identify phases with both 
ferromagnetic and anti-ferromagnetic orderings. We also find periodically modulated 
orderings formed by a cluster of like-spins followed by another cluster of opposite like-spins.
The quantum phases in the model are found to be separated by either first or second 
order transition lines.

\end{abstract}


\maketitle

\section {\label{sec:I} Introduction}

There has recently been considerable effort to understand magnetic phase 
transitions in quantum systems described by Hamiltonians with multi-spin interactions.
Ultra cold atoms trapped in optical lattices under idealized laboratory 
conditions  in particular are suitable to simulate these systems ~\cite{Sim11,Jac98,Kas95,Gre02,Man03,Pac04,Pac04b}. 
A variety of  spin Hamiltonians  have been physically realized on optical lattices, making  
possible the experimental study of  the zero temperature phase diagrams of  
those systems.  

Recently,   Simon at al.~\cite{Sim11} presented a detailed  procedure for the 
experimental realization of Ising anti-ferromagnetic spin chains in the presence of 
longitudinal and transverse magnetic  fields. Their work opened  new possibilities for 
the investigation of quantum magnetism and criticality in these systems. 

Besides the usual competition  between  various magnetic ordered ground-states, 
such as ferromagnets, antiferromagnets and paramagnets, the advent of optical 
lattices allow the study of more complex interactions that give rise to novel 
ground-state properties in magnetic systems~\cite{Pac04b,DCr04}. 
The presence of  higher order spin interactions usually  induces unusual properties 
not found in regular spin systems, bringing out a richer criticality. 

Theoretical investigations of quantum phase transitions in magnetic spin chains with  
three-  and four-spin  exchange  interactions have revealed novel phases and 
ground-states with multiple periodic structures and unique entanglement 
properties~\cite{DCr04,Pen84,Bon06,Bon13}. 
In particular, the influence of magnetic fields on low-dimensional quantum spin 
systems with complex interactions is a subject of great interest that may lead to 
the observation of reentrant behaviors and high-field driven transitions~\cite{McC11,Que12}.  

The remarkable success of the  experimental work in optical lattices simulating these 
spin systems has contributed significantly to the  renewed interest in the 
theoretical study  of these  quantum models. 
To our knowledge,  the effects of an additional longitudinal magnetic field on 
the ground-state properties of the four-spin quantum chain in a transverse magnetic 
field has never been investigated and is the subject of this paper. 
Our aim is to obtain the phase diagrams and to understand the nature of the 
phase transitions and ground-state properties of the model.

\section {\label{sec:II} The Model}

Consider a spin-1/2 magnetic chain with periodic boundary conditions. 
The spins are subjected to a magnetic field with components in the longitudinal and 
transverse directions. 
The interaction among the spins is dictated by a four-spin Ising-like term. 
The Hamiltonian of the system may be written as:
\begin{equation}
{\cal H} = - J_{4} \sum_{i}\sigma^z_{i}\sigma^z_{i+1}\sigma^z_{i+2}\sigma^z_{i+3}  
- H_x \sum_{i} \sigma^x_{i} -  H_z \sum_{i}\sigma^z_{i}.
\label{eq:Hamilt1}
\end{equation}
Here,  $\sigma^{\alpha}_i$ ($\alpha =x,y,z$) are the components of the Pauli operator, 
located at site $i$. 
The parameter $J_{4}$ is the Ising-like four-spin interaction strength. The uniform magnetic field has
components $H_{\rm x}$ and $H_{\rm z}$ along the transverse and longitudinal directions, respectively.

For $H_{\rm x}= 0$ quantum fluctuations are absent, however, depending on the
values of fields and couplings,the model may show
a variety of phases.  For instance, when the four-spin coupling $J_4 > 0.0$,
the sign of the longitudinal field $H_{\rm z}$ determines the
direction of the magnetization.  
For $H_{\rm z} > 0$, the system shows a classical ferromagnetic phase with all the spins
aligned in the $+ z$-direction, the F($+z$) phase.  
On the other hand, if $H_{\rm z} <0$, the ensuing
phase has net magnetization along the $-z$-direction, the  F($z$) phase.  

The case where  $J_4 < 0$ shows four 
phases, namely the ferromagnetic  F($\pm z$) and   $<$3,1$>$$(\pm)$ phases.  
The latter are formed by three consecutive up (down) spins followed by
one down (up) spin. 
There is a transition point at $(H_z, H_x) = (0.0, 0.0)$ between
the  $<$3,1$>$$(+)$ and $<$3,1$>$$(-)$  phases, as well as at 
$H_{\rm z},H_{\rm x})= (\pm 4.0,  0.0)$, separating the $<$3,1$>$$(+)$ from the F(+$z$)
and the   $<$3,1$>$$(-)$ from the F($-z$) phases.
The particular case of the transverse four-spin Ising model ($H_{\rm x} \ne 0$ and $H_{\rm z} = 0$)
was shown to be self-dual, with critical points at $J_4/H_{\rm x}= \pm 1$ \cite{Tur82,Pen82}.

In this paper we investigate  the ground-state properties of the Hamiltonian 
model (Eq.~\ref{eq:Hamilt1}) by using two numerical methods: entanglement entropy 
and finite-size scaling.
The first method  is based on the  behavior of 
von Neumann entanglement entropy, which is mostly used in information theory.  
That method enables one to calculate the location of the quantum critical points with a relatively 
high degree of accuracy, as well as it provides a way to identify the nature of the transitions. 
In addition, the method makes it possible to determine the  central charge of the associated conformal field 
theory with low computational cost,  by using small lattice sizes~\cite{Xav11,Nish11}. 
 The second method is based on finite-size scaling arguments, which can be used 
to determine the  transition lines and  global properties  of the various 
ground-states~\cite{Gui02}.

\section{\label{sec:III} The Methods}

\subsection{\label{A} Entropy entanglement}

In this section we describe the entropy entanglement method and show how to use it 
to locate the boundary between quantum phases, and how to find the central 
charge of the associated conformal field theory.

Consider a spin chain of length  $L$ that can be partitioned into two subsystems 
$\cal A$ and $\cal B$ of sizes $L_{\cal A} = l$ and $L_{\cal B} = L - l$, respectively. 
When the entire system is in a pure state  $|\psi\rangle$, its entropy is zero.  
However, the entropy of each subsystem is finite and can be 
quantified by the von Neumann entropy, defined as:
\begin{equation}
{S}(L,l)=-\textrm{Tr} \left( \rho_{\cal A} \ln \rho_{\cal A} \right) = -\textrm{Tr} \left( \rho_{\cal B} \ln \rho_{\cal B} \right), 
\label{vn}
\end{equation}
where $\rho_{\cal A(\cal B)}=\textrm{Tr}_{\cal B(\cal A)}\, \rho$ denotes the reduced 
density matrix of $\cal A(\cal B)$ and  $\rho=|\psi\rangle\langle \psi|$ is the density matrix of the pure state. 
The von Neumann entropy $S(L,l)$ gives a reliable measure of the entanglement between the
subsystem $\cal A$ and the rest of the system $\cal B$. 

For finite systems,  Calabrese and Cardy~\cite{Cal04}  showed that conformal invariance implies 
a diverging logarithmic scaling for the entanglement entropy. 
In particular, for a one dimensional system of size $L$ with imposed periodic boundary conditions, 
it assumes the form: 
\begin{equation}
{S}(L,l) = \frac{c}{3} \ln[\frac{L}{\pi}\sin(\frac{\pi l}{L})] + \beta,
\label{eq:vns}
\end{equation}
where $c$ is the central charge of the underlying conformal field theory and $\beta$ is a non-universal
constant which depends on the model being used.

To locate the boundary between possible quantum phases, we first calculate the entanglement entropy
difference between two subsystems with sizes $l$ and $l'$~\cite{Xav11,Nish11}:
\begin{equation}
\Delta {S} = {S}(L,l) - {S}(L,l^{\prime}) 
\label{eq:deltaS}
\end{equation}
 where $L$ is the size of the spin system.

Consider initially a system that undergoes a second-order phase transition 
when a parameter $\lambda$ of its Hamiltonian reaches a critical value $\lambda_c$.
If the system is finite,  the entanglement entropy difference remains finite 
for all values of $\lambda$, reaching a maximum at $\lambda_c$. 
As the size of the system $L$ is increased, the peak of $\Delta S$ at $\lambda_c$  
becomes progressively  narrower. Its value at the transition tends to a finite value,
whereas its value elsewhere tends to zero.

Next consider the case of a system that undergoes a first-order transition. 
Although $\Delta S$ still shows a maximum at the transition point, 
it diminishes everywhere as $L \rightarrow \infty$. 
Such behavior of the entanglement entropy difference is used as an indicator 
of the boundary between two phases and to identify the nature of the transition at that point.

In the scaling regime, where Eq.~\ref{eq:vns} is valid, we have $1 <<l,l' << L$. 
As a practical matter, to fulfill these conditions and minimize finite-size effects, 
we choose  $l= L/2$ and $l' = L/4$ in our calculations~\cite{Xav11}. Using these values
for the subsystems sizes and Eqs. \ref{eq:vns} and \ref{eq:deltaS}, we obtain:
 \begin{equation}
c = 6\,{\Delta S}/\ln(2)\,.
\label{charge}
\end{equation}
A systematic increase of the system size $L$ and the subsequent extrapolation 
to the infinite-size limit will provide an estimation of the value of the central charge.

\subsection{\label{B} Finite-size scaling}

The finite-size scaling method is another way to  
locate the boundaries between different quantum phases. This method requires
knowledge of the first two lowest energy states of the Hamiltonian, $E_0$ and $E_1$. 

Consider again a Hamiltonian model that depends 
on a parameter $\lambda$ that becomes critical at $\lambda_c$. 
It has been pointed out~\cite{Bar93} that for a system undergoing a second-order phase 
transition, the energy gap between
the two lowest energy states of the system, 
$ G(\lambda) = E_{1}(\lambda) - E_{0}(\lambda)$,
vanishes at the infinite-size limit. 
For a finite system, at criticality it obeys the following power-law dependence 
with the size $L$ of the system: 
\begin{equation}
G(L,\lambda_c) \equiv [E_{1}(L,\lambda_c) - E_{0}(L,\lambda_c)] \propto L^{-z}.   
\label{eq:a1}
\end{equation}
Here $z$ represents the dynamical critical exponent of the system~\cite{Bar93} 
which, for one-dimensional systems that are conformal invariant,  equals to one. 
For simplicity, from now on we set $z=1$ in all expressions in which it appears.

The finite-size estimation of the critical parameter $\lambda_c(L,L')$ is dependent on the choice 
of the two system sizes $L$ and $L'$. The critical point is then found as a solution of the 
phenomenological renormalization equation: 
\begin{equation}
L G(L,\lambda_c)= {L'}G(L',\lambda_c).
\label{eq:a2}
\end{equation}
The infinite-size value of the critical parameter is calculated by extrapolating the values obtained
 from Eq.~\ref{eq:a2} using increasingly larger system sizes   $L$ and $L'$.

 The ground-state and the first excited state energies and  their corresponding eigenstates are 
 calculated as a function of $\lambda$ by using a modified Lanczos method~\cite{Dag85}. 
 To speed up the calculations we use trial initial vectors which are as close as possible to the actual 
 ground state vectors. The eigenvectors and eigenvalues for the ground-states are
 determined with precision between $10^{-10}$ and $10^{-12}$.  The same 
 quantities for the first excited states are obtained with precision  between $10^{-5}$ to $10^{-6}$.

To identify the nature of each phase we need to examine the 
 corresponding ground-state eigenvectors. We start by writing the Hamiltonian on a basis 
 that consists of the product of the eigenstates  $|s>_i$ ($s= 0, 1$) of the spin operator 
 $S_i^z$, $i = 1, \dots, L$. Here the labels $s=0$ and $s=1$ correspond 
to the z-component of the spin state at the site $i$, pointing down and up respectively. Now,
an arbitrary basis state of the full Hamiltonian can be written as $|n> = \prod_i^L\, |s>_i$,  with 
the basis state labels $n = 0, 1,..., N-1$, where $N = 2^L$ determines the dimension of the
Hilbert space for a given system size $L$. An arbitrary state of the system can now be written as:
\begin{equation}
|\psi_{\alpha}> = \sum_{n=0}^{N-1}b_{\alpha}(n)|n>,
\label{eq:psi}
\end{equation}
 where $\alpha = 0$ labels  the ground-state, and $\alpha=1$ the first excited state. 
 The coefficients $b_{\alpha}(n)$ are the amplitudes of each of the basis states $|n$$>$, of
 the linear combination forming the arbitrary state $|\psi_{\alpha}>$. 
Those coefficients are all real, since the Hamiltonian 
 matrix is real and symmetric.

\begin{figure}
\includegraphics[width=8.0cm, height= 6.0cm, angle=0]{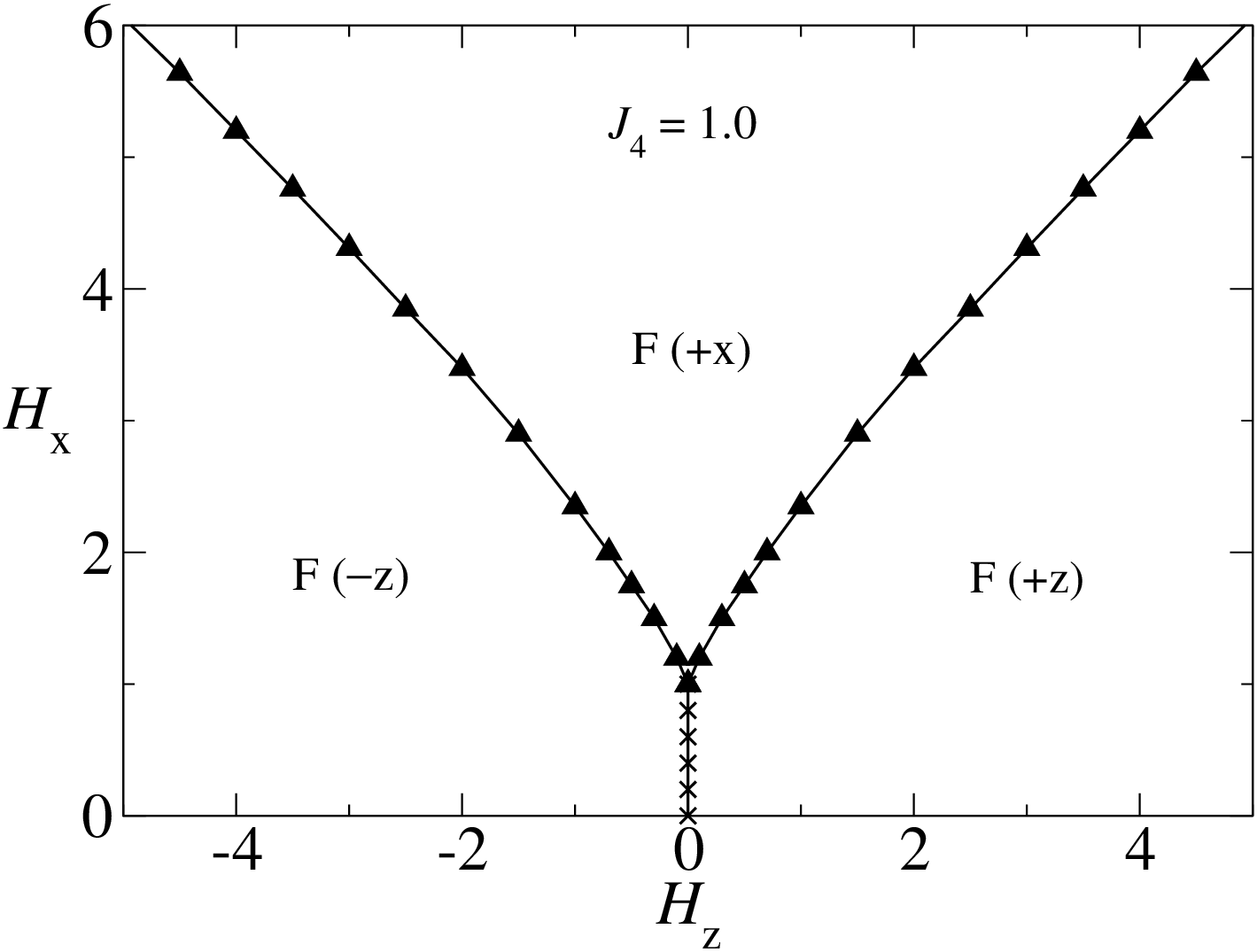}
\setlength{\abovecaptionskip}{0pt}
\setlength{\belowcaptionskip}{10pt}
\caption{Phase diagram in the $(H_{\rm z}$-$H_{\rm x}$)-plane for  $J_4=1.0$. 
The triangles separate a field induced ferromagnetic phase with net 
magnetization in the $x$-direction, F($+x$),  from two induced phases  
with magnetizations along the $\pm z$-directions, F(+$z$) and F($-z$).  The crosses 
separate the ferromagnetic phases along $\pm z$.  All the
transition lines are of first order.  The data shown were obtained
with the entanglement entropy method.
\label{fig:fig1}
}
\end{figure}

The basis state labels $n$, can be written in binary notation with $L$ digits.
The $i$-th position and the value of these digits coincide with the eigenstate 
of ${S_i^z}$ at that site. 
 By plotting the coefficients  $b_{\alpha}(n)$ as a function of the basis label $n$, 
we obtain a  representation of the quantum state on a single graph and a full characterization 
of the nature of that state~\cite{Bon06}.

\section{\label{sec:IV} Results}

We have carried out numerical calculations to investigate the quantum phase 
transitions of the Hamiltonian, Eq.~\ref{eq:Hamilt1}, using 
the methods of entanglement entropy and finite-size scaling.  
We considered chains containing up to 24 spins, and used periodic boundary conditions.

First we set $J_4=1.0$  and search for phase 
transitions by varying the magnetic field  
components $H_{\rm z}$ and $H_{\rm x}$.  
Our numerical results for the phase diagram in the ($H_z$-$H_x$)-plane are shown  
in Fig.~\ref{fig:fig1}.  
The transition lines (with triangles) separate a ferromagnetic phase 
with net magnetization in the $x$-direction F($+x$) from
two ferromagnetic phases F($+z$) and F($-z$) with spins aligned along the $+z$ and $-z$ directions, respectively.

In the absence of the four-spin interaction, the phase transition lines in Fig.~\ref{fig:fig1}
 would be along the lines $H_{\rm x} = \pm H_{\rm z}$. Under the present 
conditions however, the field $H_z$ reinforces the ferromagnetic order caused by the four-spin 
interaction.  Therefore, it takes a larger transverse field $H_x$ to change the
direction of the net magnetization from the $\pm z$- to the $x$-direction
as $H_z$ increases.

Notice that at $H_z=0.0$, the critical transverse field is given by  $H_x = J_4 = 1.0$,
a known result~\cite{Tur82,Pen82}. 
For $H_{\rm z }= 0$ and $ 0.0 < H_{\rm x} < 1.0$, there is a transition line (with crosses) separating the
ferromagnetic phases F($+z$) and F($-z$).
Along that line the quantum state is predominantly formed by states with 
ferromagnetic, anti-ferromagnetic and $<$$2$,$2$$>$ orderings. 
The latter is a modulated ordering formed by two up spins followed by two down spins or vice-versa.

\begin{figure}
\includegraphics[width=8.0cm, height= 6.0cm, angle=0]{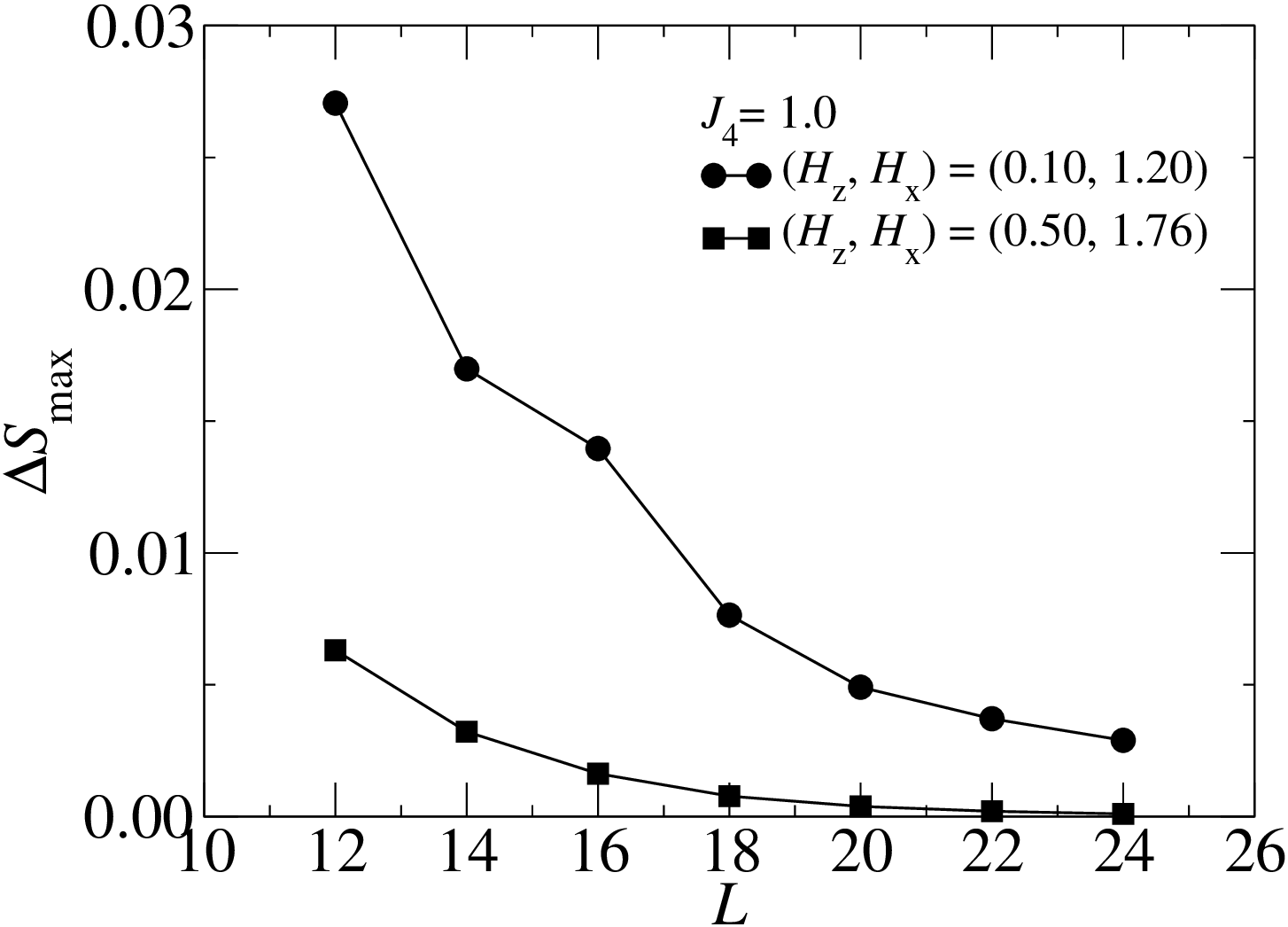}
\setlength{\abovecaptionskip}{0pt}
\setlength{\belowcaptionskip}{10pt}
\caption{Peak values of the entanglement entropy difference $\Delta {S_{\max}}$, as 
a function of the size of the system $L$, for $J_4 = 1.0$ 
and critical fields  $(H_{\rm z}, H_{\rm x}) $= ($0.10 , 1.20$) and ($0.50, 1.76$), 
along the transition line between the ferromagnetic phases in the +$x$ and +$z$ 
directions, which is shown in Fig.~\ref{fig:fig1}. 
In the two cases depicted, $\Delta {S_{\max}}$ decreases with $L$, indicating that  the
transition between the field induced phases is of first order.
\label{fig:fig2}
}
\end{figure}

The nature of the phase transitions is inferred from the dependence
of the maximum of the entanglement entropy difference $\Delta S_{\rm max}$ 
as a function of the system size $L$. 
Figure~\ref{fig:fig2} shows the results for the case $J_4 = 1.0$ and
critical fields  $(H_{\rm z}, H_{\rm x}) $= ($0.10 , 1.20$) and ($0.50, 1.76$).
The data points were obtained along the transition line between the ferromagnetic  phases 
in the +$x$ and +$z$ directions, which is shown in Fig.~\ref{fig:fig1}.
$\Delta S_{\rm max}$ decreases with $L$, suggesting that 
the transition is of first order. 
The other transition lines of Fig.~\ref{fig:fig1} produce similar behavior 
for $\Delta S_{\rm max}$, indicating that the
transitions are all of first order.

\begin{figure}
\includegraphics[width=8.0cm, height= 6.0cm, angle=0]{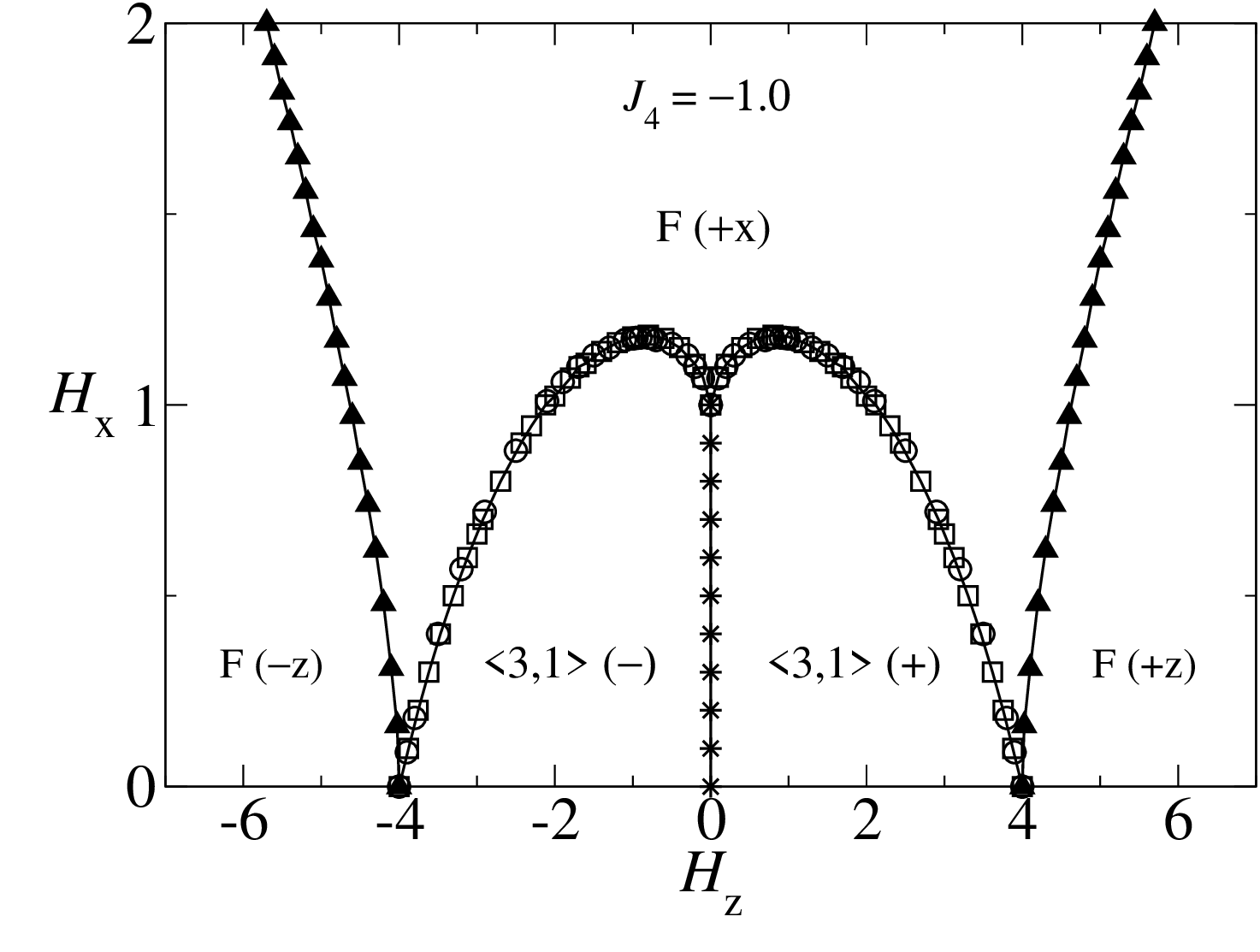}
\setlength{\abovecaptionskip}{0pt}
\setlength{\belowcaptionskip}{10pt}
\caption{ Phase diagram in the ($H_{\rm z}$-$H_{\rm x}$)-plane for  the case $J_4=-1.0$. 
The squares and circles lie on a second-order transition line which separate 
a ferromagnetic phase with net magnetization in the
$x$-direction F(+$x$) from the phases $<$3,1$>$$(+)$, for $H_z>0$ and
$<$3,1$>$$(-)$, for $H_{\rm z}<0$.  
For $H_{\rm z} = 0$ and $0 \le H_{\rm x} \le 1.0$, there is
a first-order transition line (with stars) between the  $<$3,1$>$ phases.
For $|H_{\rm z}| > 4$ there are two  first-order transition lines (with triangles) separating
the ferromagnetic phase F($+x$) from the ferromagnetic
phases F($+z$) and F($-z$).
The squares were obtained using finite-size scaling,
while the other data points were determined by 
entanglement entropy. 
 \label{fig:fig3}
}
\end{figure}

By reversing the sign of the four-spin interaction to $J_4 = -1.0$, the model shows 
a richer phase diagram in the $(H_{\rm z}$-$H_{\rm x}$)-plane, 
which is shown in Fig.~\ref{fig:fig3}.  For low fields, the phases are
the $<$3,1$>$ ground-states, together with background noise-like components 
caused by the transverse field. 
For $H_{\rm z} > 0.0$, the ground-states
are dominated by the sequence of three spins up followed by one spin down,
the $<$3,1$>$$(+)$ phase.  Conversely, for  $H_{\rm z} < 0.0$ the
ground-state consists of the sequence of three spins down followed by one spin up,
the $<$3,1$>$$(-)$ phase.
There is a first-order transition line (with stars) between these two phases 
along the line segment  $0.0 \le H_{\rm x} \le 1.0$ located at $H_{\rm z} = 0.0$. 
There, the quantum state with most dominant components exhibit
both $<$3,1$>$$(+)$ and $<$3,1$>$$(-)$ orderings.
In the region $|H_{\rm z}| < 4.0$, there are two second-order
transition lines (with squares and circles) separating the F($+x$) phase
from the $<$3,1$>$ phases.  
These lines merge at the multi-critical point
($H_{\rm z}$, $H_{\rm x}$)$=$($0.0$, $1.0$).
There are two other multi-critical points, located at 
 $(H_{\rm z}, H_{\rm x})$=$(\pm 4.0, 0.0)$,
where first- and second-order transition lines meet. 
For $|H_{\rm z}| \ge 4.0$, there are two regions 
of ferromagnetic phases,  F($+z$) and F($-z$),
where the spins are mostly aligned along the $+z$ or $-z$ directions. 
As $H_{\rm x}$ increases, the competition between these
phases and the  F($+x$) phase produces phase transition lines
of first order (with triangles).   

\begin{figure}
\includegraphics[width=8.0cm, height= 6.0cm, angle=0]{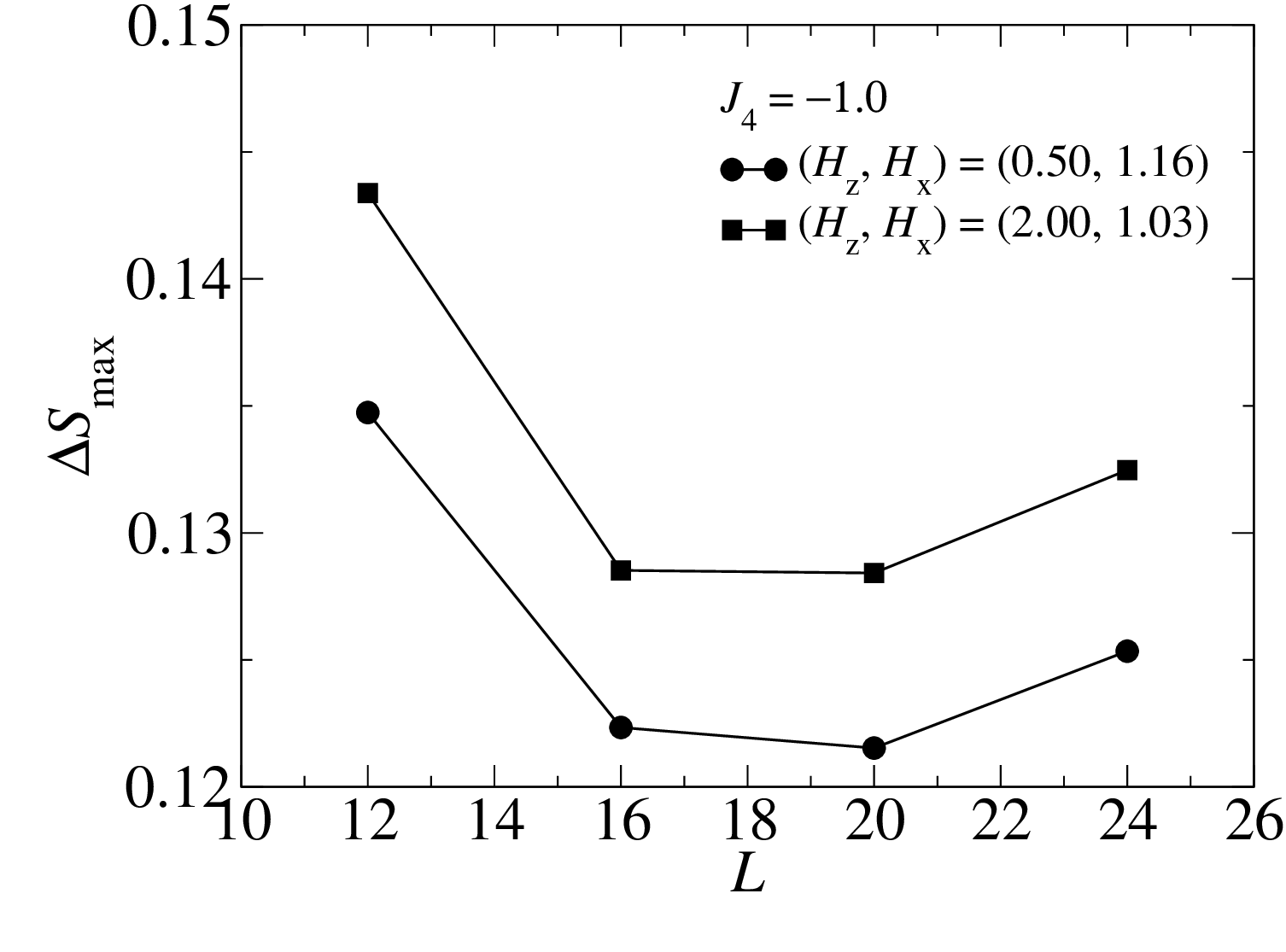}
\setlength{\abovecaptionskip}{0pt}
\setlength{\belowcaptionskip}{10pt}
\caption{
Maximum of the entanglement entropy difference $\Delta {S_{\max}}$ 
{\em vs}  system size $L$ 
for two values of the critical fields  $(H_{\rm z}, H_{\rm x})$ = $(0.50, 1.16)$ 
and $(2.00, 1.03)$, along the transition line in the region $0 \le H_{\rm z} \le 4$ 
of Fig.~\ref{fig:fig3}. 
The increase of $\Delta {S_{\max}}$ with $L$ indicates that 
the phase transition is of second order.  
\label{fig:fig4}
}
\end{figure}

The numerical analysis leading to the nature of the transitions is again
based on the behavior of the maximum of the entanglement
entropy difference versus the system size.  
The entropy differences along the transition lines between the
 F($+x$)  and  F($\pm z$) show similar behavior as those shown in
Fig.~\ref{fig:fig2}, therefore they can be viewed as first-order
transition lines.  
On the other hand, the transition lines between 
the   $<$3,1$>$$(\pm)$ and the F($\pm z$) phases 
can only be analyzed in lattices
with periodicity of 4 site spacings.  That is, the size $L$ must
be a multiple of 4, so as to make the lattice commensurate with
the   $<$3,1$>$$(\pm)$ orderings.  
Figure~\ref{fig:fig4} shows
the behavior of the maximum of the entanglement entropy difference
$\Delta {S_{\max}}$ for $L = 12, 16, 20,$ and $24$.  The data were
obtained along the transition line in the region 
$0.0 \le H_{\rm z} \le 4.0$ of   Fig.~\ref{fig:fig3}
for two values of the critical fields,   
$(H_{\rm z}, H_{\rm x})$ = $(0.50, 1.16)$ and $(2.00, 1.03)$.
At first, $\Delta {S_{\max}}$ decreases with $L$.  Then it passes
through a a minimum and rises between $L=20$ and $24$. 
We believe this trend will continue, so that the transition 
is of second order.  
Unfortunately,  at present, it is  numerically prohibitive to tackle larger lattices,
considering that the next relevant size would be $L=28$.

\begin{figure}
\includegraphics[width=8.0cm, height= 6.0cm, angle=0]{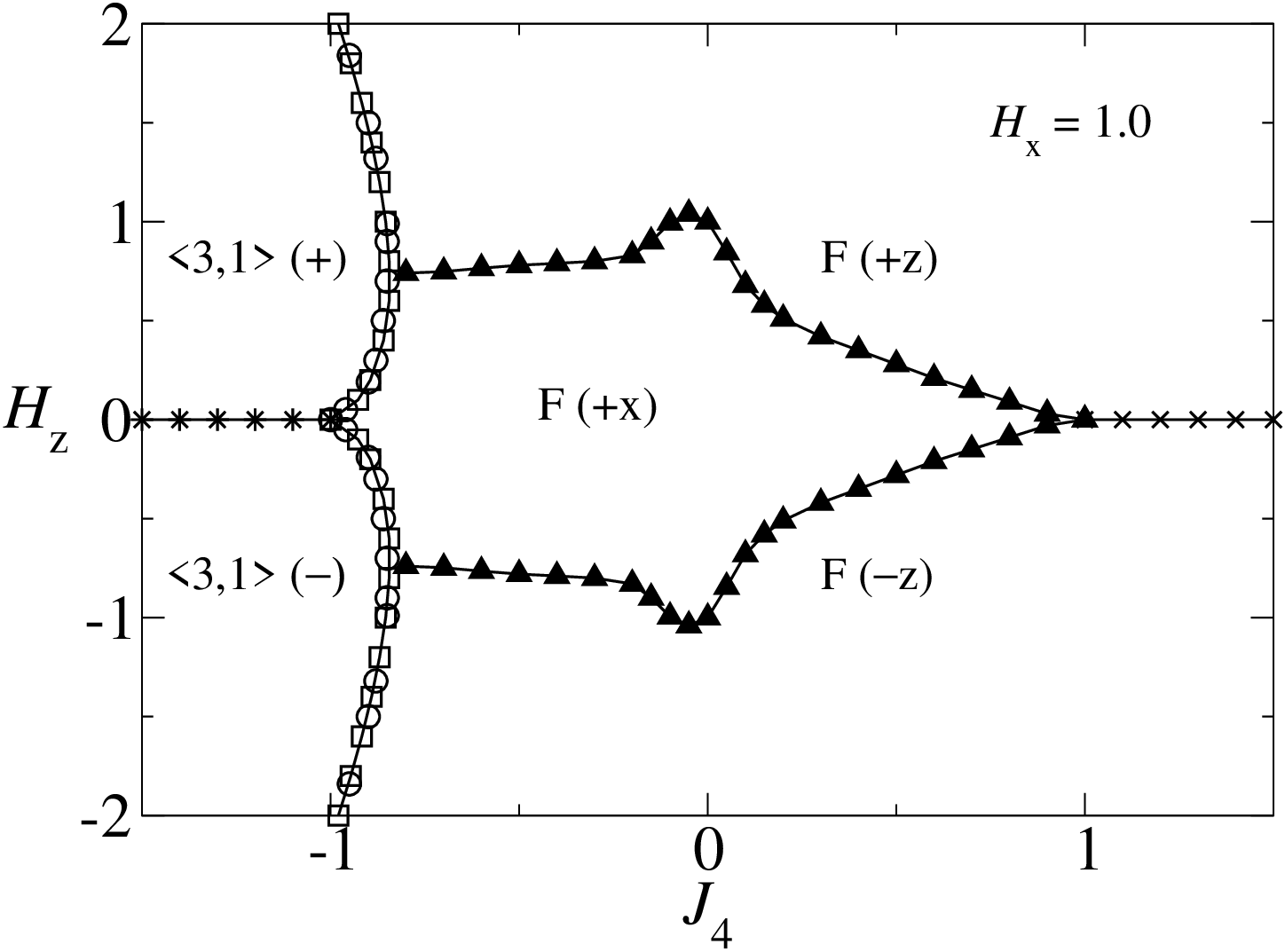}
\setlength{\abovecaptionskip}{0pt}
\setlength{\belowcaptionskip}{10pt}
\caption{Phase diagram in the ($J_4$-$H_{\rm z}$)-plane for the case $H_{\rm x}$=$1.0$. 
The triangles separate a field-induced ferromagnetic phase with net magnetization in the +$x$ direction,
F(+$x$), from two other ferromagnetic phases with magnetizations along the $\pm z$ directions,
F(+$z$) and F($-z$). 
The phases   $<$3,1$>$$(+)$ and $<$3,1$>$$(-)$ are separated by
a first-order transition line (with stars).
For $J_4 > 1$, there is a first-order transition line (with crosses)
separating the two field-induced ferromagnetic phases
F(+$z$) and F($-z$).
The squares are results from finite-size scaling, while the
other data are from entanglement entropy.
 \label{fig:fig5}
}
\end{figure}

For completeness we also  perform  calculations for $H_{\rm x} =1.0$ 
to investigate the occurrence of phase transitions when the transverse 
field is kept constant, so that the system is always in the quantum regime.
The phase diagram in the  ($J_4$-$H_{\rm z}$)-plane  is shown in  Fig.~\ref{fig:fig5}.
The ferromagnetic phase with net magnetization in the $x$-direction F($+x$) appears
as an island surrounded by the $<$3,1$>$$(+)$,  $<$3,1$>$$(-)$, F($+z$), and F($-z$)
phases.  

The transition lines separating the $<$3,1$>$  from the F  phases 
are of second order.
The other transition lines are all of first order.
Along the boundary line separating the $<$3,1$>$ phases, the
ground-states are quantum states containing equal contributions from 
$<$3,1$>$$(+)$ and $<$3,1$>$$(-)$ configurations, as well as from a background
of states induced by the transverse field.
On the other side of the diagram for $J_4 > 1.0$, along the transition line between
the two F phases,  the ground states are formed by the coexistence of 
ferromagnetic, antiferromagnetic, and $<$2,2$>$ orderings, with additional background
states caused by the presence of the transverse magnetic field.
Finally, there are four multi-critical points, which are located at 
($J_4$, $H_{\rm z}$)$=$($-0.84 \pm 0.01$, 
$\pm 0.74 \pm 0.01$) and ($ \pm 1.0$, $0.0$).

\section{\label{sec:V} Conclusions}

We have  studied the competing effects of a magnetic field with components in the longitudinal and 
transverse directions on the quantum behavior of an Ising-like chain with four-spin interactions.  
The entanglement entropy and finite-size scaling methods have been used to obtain the phase diagrams of the system. 
A rich variety of quantum phases and multi-critical points have been show to be present in the model. Both 
first and second order transitions are observed among the phases. Under certain conditions the physics of 
atoms interacting in a one dimensional lattice may be captured by the model Hamiltonian analyzed here. 
It would be interesting to see how optical lattice techniques could be implemented to simulate the
present model.

We thank FAPERJ (Brazilian agency)  and PROPPI/UFF for financial support.
(O.F.A.B.) acknowledges support from the Murdoch
College of Science Research Program and a grant from the 
Research Corporation through the Cottrell College Science 
Award No.~CC5737.



\begin{thebibliography}{}


\bibitem{Sim11} J.~Simon, W.S.~Bakr, R.~Ma, M.E.~Tai, P.M.~Preiss, and M.~Greiner, Nature (London) {\bf 472}, 307 (2011).

\bibitem{Jac98} D.~Jacksch, C.~Bruder, J.I.~Cirac, C.W.~Gardiner, and P.~Zoller, Phys. Rev. Lett. {\bf
81}, 3108 (1998).

\bibitem{Kas95} A.~Kastberg, W.D.~Phillips, S.L.~Rolston, R.J.C.~Spreeuw, and P.S.~Jessen, Phys. Rev. Lett. {\bf
74}, 1542 (1995); G.~Raithel, W.D.~Philips, and S.L.~Rolston, {\it ibid}. {\bf81}, 3615 (1998).

\bibitem{Gre02} M.~Greiner, O.~Mandel, T.~Esslinger, T.W.~Heanch, and I.~Bloch, Nature (London) {\bf 415}, 39 (2002);
M.~Greiner, O.~Mandel, T.W.~Heanch, and I.~Bloch, {\it ibid} {\bf 419}, 51 (2002).

\bibitem{Man03} O.~Mandel, M.~Greiner, A.~Widera, T.~Rom, T.W.~Heanch, and I.~Bloch, Nature (London) {\bf 425}, 937 (2003).

\bibitem{Pac04} J.K.~Pachos and E.~Rico, Phys. Rev. A {\bf 70}, 053620 (2004).

\bibitem{Pac04b} J.K.~Pachos and M.B.~Plenio, Phys. Rev. Lett. {\bf 93}, 056402 (2004).

\bibitem{DCr04} C.~D'Cruz and J.K.~Pachos, Phys. Rev. A {\bf 72}, 043608 (2005).

\bibitem{Pen84} K.A~Penson, Phys. Rev. B {\bf 29}, 2404 (1984).

\bibitem{Bon06} O.F.~de~Alcantara~Bonfim and J.~Florencio, Phys. Rev. B {\bf 74}, 134413 (2006).

\bibitem{Bon13} O.F.~de~Alcantara~Bonfim, A.~Saguia, B.~Boechat, and J.~Florencio, to be published.

\bibitem{McC11} J.F.~McCabe and T.~Wydro, Phys. Rev. E {\bf 84}, 031123 (2011).

\bibitem{Que12} S.L.A. de Queiroz, Phys. Rev. E {\bf 84}, 031132 (2011).

\bibitem{Tur82} L.~Turban, J.~Phys.~C {\bf 15}, L65 (1982).

\bibitem{Pen82} K.A.~Penson, R.~Jullien, and P.~Pfeuty, Phys. Rev. B {\bf 26}, 6334 (1982).

\bibitem{Xav11} J.C.~Xavier, F.C.~Alcaraz, Phys. Rev. B 84, 094410 (2011).

\bibitem{Nish11} S.~Nishimoto, Phys. Rev. B 84, 195108 (2011). 

\bibitem{Gui02} P.R.C.~Guimar\~aes, J.A.~Plascak, F.C.~S\'a Barreto, and J.~Florencio, 
Phys. Rev. B {\bf 66}, 064413 (2002).

\bibitem{Cal04} P.~Calabrese, J.~Cardy, J. Stat. Mech. P06002 (2004).

\bibitem{Bar93} M.~Barber, {\em Phase Transition and Critical Phenomena\/}, Vol. {\bf 8} 
(Academic Press, New York, 1993).

\bibitem{Dag85} E.~Dagotto and A.~Moreo, Phys. Rev. D {\bf 31}, 865 (1985); 
E.R.~Gagliano, E.~Dagotto, A.~Moreo, and F.C.~Alcaraz, Phys. Rev. B {\bf 34}, 1677 (1986).





\end{thebibliography}
\end{document}